\documentclass[conference]{IEEEtran}

\IEEEoverridecommandlockouts
\usepackage{cite}
\usepackage{amsmath,amssymb,amsfonts}
\usepackage{amsthm}
\usepackage{algorithmic}
\usepackage{graphicx}
\usepackage{textcomp}
\usepackage{xcolor}
\usepackage{hyperref}
\usepackage{makecell}
\usepackage{algorithm}
\usepackage{upgreek}
\usepackage{booktabs}
\usepackage{tabularx}
\usepackage{enumitem}
\usepackage{bbold}
\usepackage{algorithmic}
\usepackage{bbm}
\usepackage{quantikz} 
\usepackage{dblfloatfix}

\newtheorem*{theorem*}{Theorem}

\hypersetup{
    colorlinks=true, 
    linkcolor=black,  
    citecolor=black, 
    filecolor=black, 
    urlcolor=black    
}

\def\BibTeX{{\rm B\kern-.05em{\sc i\kern-.025em b}\kern-.08em
    T\kern-.1667em\lower.7ex\hbox{E}\kern-.125emX}}
\begin{document}

\title{ Adaptive Resource Orchestration for\\ Distributed Quantum Computing Systems
}

\author{
\IEEEauthorblockN{
    Kuan-Cheng Chen\IEEEauthorrefmark{2}\IEEEauthorrefmark{3}\IEEEauthorrefmark{1},
    Felix Burt \IEEEauthorrefmark{2}\IEEEauthorrefmark{3},
    Nitish K. Panigrahy\IEEEauthorrefmark{4},
    Kin K. Leung\IEEEauthorrefmark{2}}

\IEEEauthorblockA{\IEEEauthorrefmark{2}Department of Electrical and Electronic Engineering, Imperial College London, London, UK}
\IEEEauthorblockA{\IEEEauthorrefmark{3}Centre for Quantum Engineering, Science and Technology (QuEST), Imperial College London, London, UK}
\IEEEauthorblockA{\IEEEauthorrefmark{4} Binghamton University SUNY, New York, USA}
\IEEEauthorblockA{\IEEEauthorrefmark{1} Email: kuan-cheng.chen17@imperial.ac.uk}
}


\maketitle

\begin{abstract}
Scaling quantum computing beyond a single device requires networking many quantum processing units (QPUs) into a coherent \emph{quantum‑HPC} system. We propose the \emph{Modular Entanglement Hub (ModEn‑Hub)} architecture: a hub‑and‑spoke photonic interconnect paired with a real‑time quantum network orchestrator. ModEn‑Hub centralises entanglement sources and shared quantum memory to deliver on‑demand, high‑fidelity Bell pairs across heterogeneous QPUs, while the control plane schedules teleportation‑based non‑local gates, launches parallel entanglement attempts, and maintains a small ebit cache. To quantify benefits, we implement a lightweight, reproducible Monte Carlo study under realistic loss and tight round budgets, comparing a naïve sequential baseline to an orchestrated policy with logarithmically scaled parallelism and opportunistic caching. Across 1–128 QPUs and 2{,}500 trials per point, ModEn‑Hub‑style orchestration sustains roughly 90\% teleportation success while the baseline degrades toward about 30\%, at the cost of higher average entanglement attempts (approximately 10–12 versus about 3). These results provide clear, high‑level evidence that adaptive resource orchestration at the ModEn‑Hub enables scalable and efficient quantum‑HPC operation on near‑term hardware.
\end{abstract}


\section{Introduction}
\label{sec:intro}

Quantum computing is poised to address problems that remain out of reach for classical machines, including cryptographic analysis, large-scale optimisation, and accurate simulation of quantum many-body systems. Contemporary quantum processing units (QPUs) nevertheless contain only tens to hundreds of physical qubits, a scale far below that required for practical quantum advantage. Many estimates suggest that thousands of logical qubits which translates to millions of physical qubits. will be needed to support error-corrected, industry-relevant workloads, a capacity beyond the foreseeable capability of any single device.

A promising route to this scale is distributed quantum computing, wherein multiple QPUs are interconnected through a quantum network and operate as a single virtual computer~\cite{cuomo2020towards}. These QPUs often differ in size, fidelity, and connectivity, and the underlying network topology may evolve over time due to link failures or reconfiguration. When processors share quantum entanglement in advance, the collective computational power can grow exponentially with the number of participating QPUs, a stark contrast to the linear scaling characteristic of classical distributed systems. Realising such an architecture is non-trivial because qubits decohere rapidly and cannot be cloned, precluding straightforward transmission of quantum states~\cite{cacciapuoti2019quantum}. Instead, remote operations rely on entangled pairs of qubits, also known as \textit{e-bits}, distributed between QPUs so that quantum teleportation and non-local gates can bind distant qubits as if they were co-located~\cite{pompili2022experimental}. Recent advances in quantum internet research illustrate steady progress: for example, Main et al.~\cite{main2025distributed} have experimentally demonstrated the distribution of quantum computations between two photonically interconnected trapped-ion modules, including the deterministic teleportation of a controlled-Z gate and the execution of distributed Grover’s search. These developments highlight the need for architectural frameworks able to orchestrate entanglement generation and management across many nodes.

We propose a Modular Entanglement Hub (ModEn‑Hub) architecture that adapts principles from classical high-performance computing to the unique constraints of quantum mechanics. This specialised hub acts as an entanglement-centric interconnect linking multiple QPU clusters. It integrates an Adaptive Entanglement Generation Module and a Quantum Network Orchestrator, coordinating the creation and distribution of entangled Bell pairs over high-fidelity photonic links—building on emerging hardware efforts focused on modular entanglement sources and quantum network interfaces. A parallel low-latency classical channel carries measurement outcomes required for feed-forward control and teleportation-based operations. By centralising entanglement sources, quantum memory, and optical switching within the hub, the architecture enables efficient resource sharing and provides dynamic, on-demand connectivity between any pair of processors, thereby overcoming the scalability limitations inherent in point-to-point or nearest-neighbour topologies.

This work makes four primary contributions. First, it introduces a modular hub-and-spoke network design that enables heterogeneous QPUs to operate coherently, effectively extending the logical qubit count beyond the physical limits of standalone devices. Second, it develops a network-aware circuit orchestration strategy that aligns quantum gate scheduling with the dynamic state of the network and coordinates classical feed-forward signals. Third, it proposes adaptive entanglement management policies, grounded in reinforcement learning techniques, that determine which links to entangle, when to schedule entanglement swaps, and how to cache Bell pairs, thereby improving request success probability. Finally, it presents a prototype-level simulator and a four-QPU case study demonstrating that the hub distributes high-fidelity entanglement with latencies compatible with near-term algorithms.

\section{Opportunities and Prospects}
\label{sec:opportunities}

\begin{figure*}[!b]
  \centering
  \includegraphics[width=0.5\textwidth]{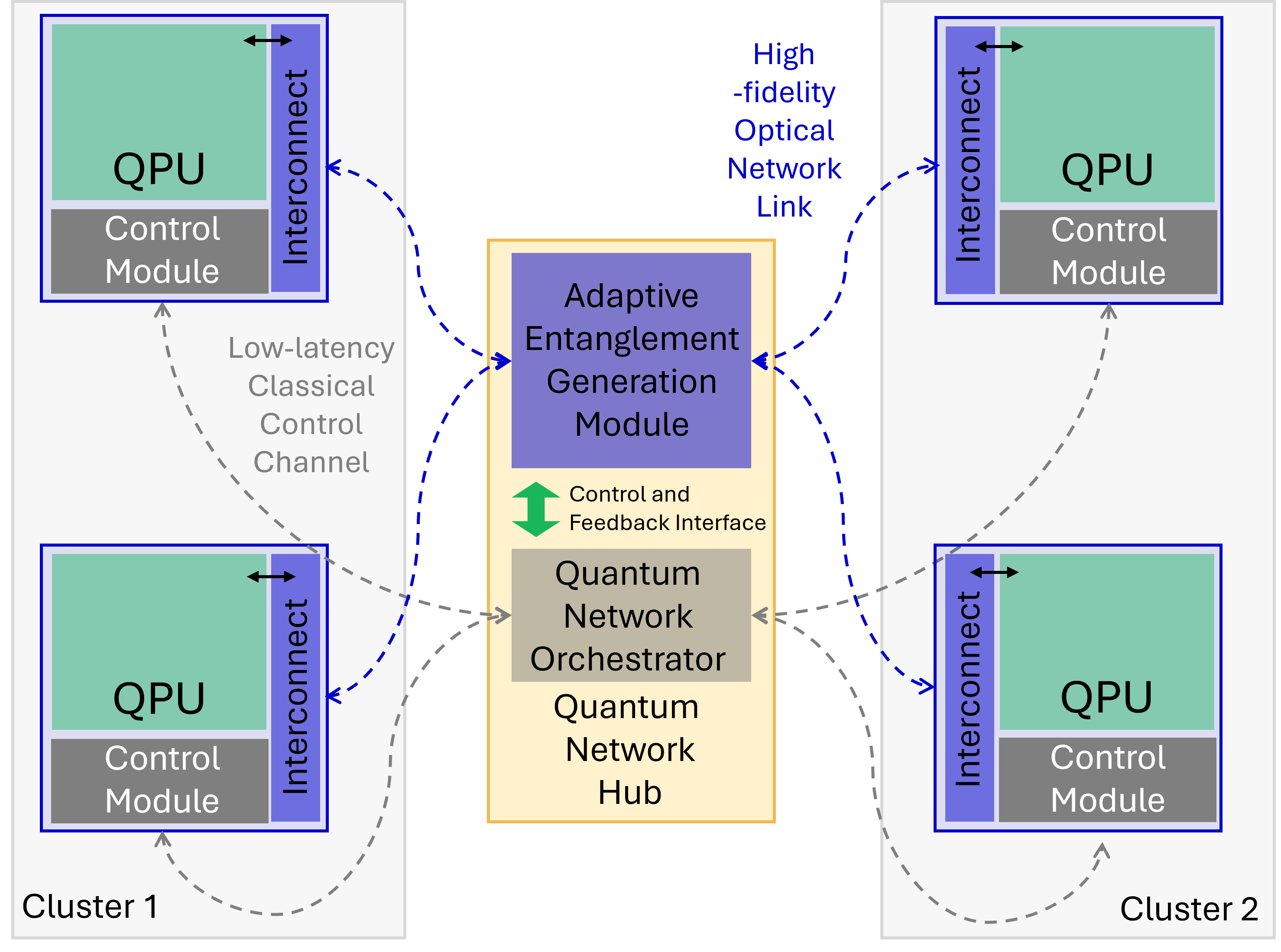}
  \caption{ModEn‑Hub Architecture for Distributed Quantum Computing.
The figure illustrates the Modular Entanglement Hub (ModEn‑Hub) architecture, which enables entanglement-based interconnectivity between two quantum computing clusters (Cluster 1 and Cluster 2). Each cluster contains one or more quantum processing units (QPUs), each coupled with a local Control Module and an Interconnect. A central Quantum Network Hub connects the clusters and comprises two main components: (1) the Adaptive Entanglement Generation Module, responsible for generating high-fidelity entangled photon pairs on demand and distributing them via optical quantum links (blue dashed lines), and (2) the Quantum Network Orchestrator, which coordinates the timing and routing of entanglement generation, quantum teleportation, and remote gate operations across QPUs. In parallel, low-latency classical control channels (gray dashed lines) carry measurement outcomes and synchronization signals between the hub and QPUs to enable real-time feed-forward control. The architecture supports scalable and heterogeneous distributed quantum computing by decoupling entanglement generation and control coordination from the quantum clusters.
}
  \label{fig:dqc-framework}
\end{figure*}

Connecting multiple processors to operate cooperatively is a time‑honoured strategy in classical computing—from cloud data centres to peta‑ and exa‑scale supercomputers.  
The \emph{quantum data‑centre} concept extends the same principle to the quantum domain: by networking several modest‑sized QPUs, one can construct a virtual machine whose qubit count and computational reach far exceed those of any individual device~\cite{cuomo2020towards}.  
In a distributed setting, an algorithm is partitioned across nodes that exchange quantum information; when executed efficiently, \(N\) small processors linked by entanglement can emulate a system with \(\gg N\) logical qubits because the joint Hilbert space grows exponentially with the number of nodes~\cite{burt2024generalised,burt2025multilevelframeworkpartitioningquantum, burt2025entanglement}. 

Partitioning requires decomposing the quantum circuit into a number of sub-circuits, which interact only via the generation of e-bits. Using graphical optimisation techniques, the required entanglement between sub-circuits can be minimised to avoid strain on the network\cite{burt2024generalised,burt2025entanglement}.

Classical networking techniques cannot be applied verbatim.  
Transmitting physical qubits—-typically photons through optical fibre—is lossy, and any loss or decoherence irreversibly destroys the quantum state~\cite{cacciapuoti2019quantum}.  
Moreover, the no‑cloning theorem forbids amplification or regeneration of unknown quantum states, ruling out classical repeater strategies.  
\emph{Entanglement} therefore becomes the essential communication resource: pairs of entangled qubits (e-bits) are distributed so that quantum‑teleportation or remote‑gate protocols can move logical states between processors without physically transporting fragile qubits~\cite{main2025distributed}.

Early architectural proposals relied on point‑to‑point links, sometimes aided by chains of quantum repeaters\cite{azuma2023quantum}. While feasible for a handful of nodes, such designs do not scale gracefully as the network size grows. Subsequent work adopted topologies inspired by classical interconnects. \emph{Switch‑centric} approaches place reconfigurable photonic switches at the core to establish ephemeral circuits among QPUs on demand, mirroring circuit‑switched telecommunication networks. \emph{Server‑centric} schemes connect each QPU to a structured subset of peers—evoking fat‑tree or hypercube layouts—to balance connectivity against hardware cost. All of these architectures must contend with the probabilistic nature of entanglement generation: photon loss, detector dark counts, and finite‑memory lifetimes mean that attempts to entangle remote qubits often fail. Consequently, considerable research targets quantum‑routing algorithms and scheduling policies that maximise entanglement throughput under latency constraints.  State‑of‑the‑art methods employ reinforcement learning to select links and to time swap operations, achieving faster decisions and higher success rates by caching Bell pairs when advantageous\cite{li2024optimising}.

An intelligent \emph{entanglement orchestrator} has thus emerged as a critical control‑plane component.  
It dynamically reconfigures connectivity between QPUs by instructing local operations on distributed entangled states and by allocating limited entanglement resources to meet algorithmic demands~\cite{d2025modelling}.  
Recent prototypes integrate tunable entanglers with real‑time orchestration logic and have demonstrated multi‑node links with high fidelity, indicating that the underlying control‑plane concepts are technologically viable.  
Major platform providers are likewise exploring middleware that decomposes quantum circuits across multiple processors and later recombines the results\cite{mengoni2025efficient}.

In summary, the literature identifies two foundational pillars for scalable distributed quantum computing: (i) \textbf{quantum-network hardware}, including entanglement sources, quantum memories, optical switches, and transducers; and (ii) \textbf{quantum-network control}, encompassing algorithms for routing, scheduling, and orchestration. The framework presented in this paper integrates these data-plane and control-plane components into a unified system designed for practical quantum–data-centre deployment. Table~\ref{tab:opportunities} distils the principal opportunities enabled by adaptive resource orchestration.

\begin{table}[t]
  \centering
  \caption{Opportunities enabled by adaptive resource orchestration in distributed quantum computing.}
  \label{tab:opportunities}
  \renewcommand{\arraystretch}{1.1}
  \begin{tabular}{p{0.28\linewidth}p{0.65\linewidth}}
    \toprule
    \textbf{Opportunity} & \textbf{Implication for DQC} \\
    \midrule
    Virtual large‑scale QPU &
      Entangling \(N\) modest devices yields an effective Hilbert‑space dimension \(2^{\sum_i n_i}\), surpassing the qubit count of any single node. \\[0.3em]

    Dynamic topology adaptation &
      Control‑plane decisions re‑route entanglement around lossy links, preserving connectivity as physical conditions drift. \\[0.3em]

    Resource‑aware scheduling &
      Joint optimisation of link usage and local‑gate queues reduces idle time and maximises algorithmic throughput under decoherence constraints. \\[0.3em]

    Load balancing &
      Temporal multiplexing of Bell‑pair generation equalises utilisation across heterogeneous hardware, mitigating bottlenecks. \\[0.3em]

    Fault tolerance &
      Adaptive allocation of e-bits to error‑corrected logical qubits supports distributed surface codes\cite{kim2024fault} and enhances resilience to local failures. \\[0.3em]

    Hybrid cloud integration &
      Orchestration middleware exposes a unified API, enabling seamless inclusion of remote QPUs as accelerators within classical workflows. \\[0.3em]

    Algorithmic co‑design &
      Real‑time knowledge of network state informs circuit partitioning and compilation heuristics, lowering total entanglement cost for VQE, QAOA, and related algorithms\cite{burt2025entanglement}. \\
    \bottomrule
  \end{tabular}
\end{table}

\section{Architectural Framework: A Quantum Network Hub Connecting QPU Clusters}

\begin{figure*}[!b]
  \centering
  \includegraphics[width=0.63\textwidth]{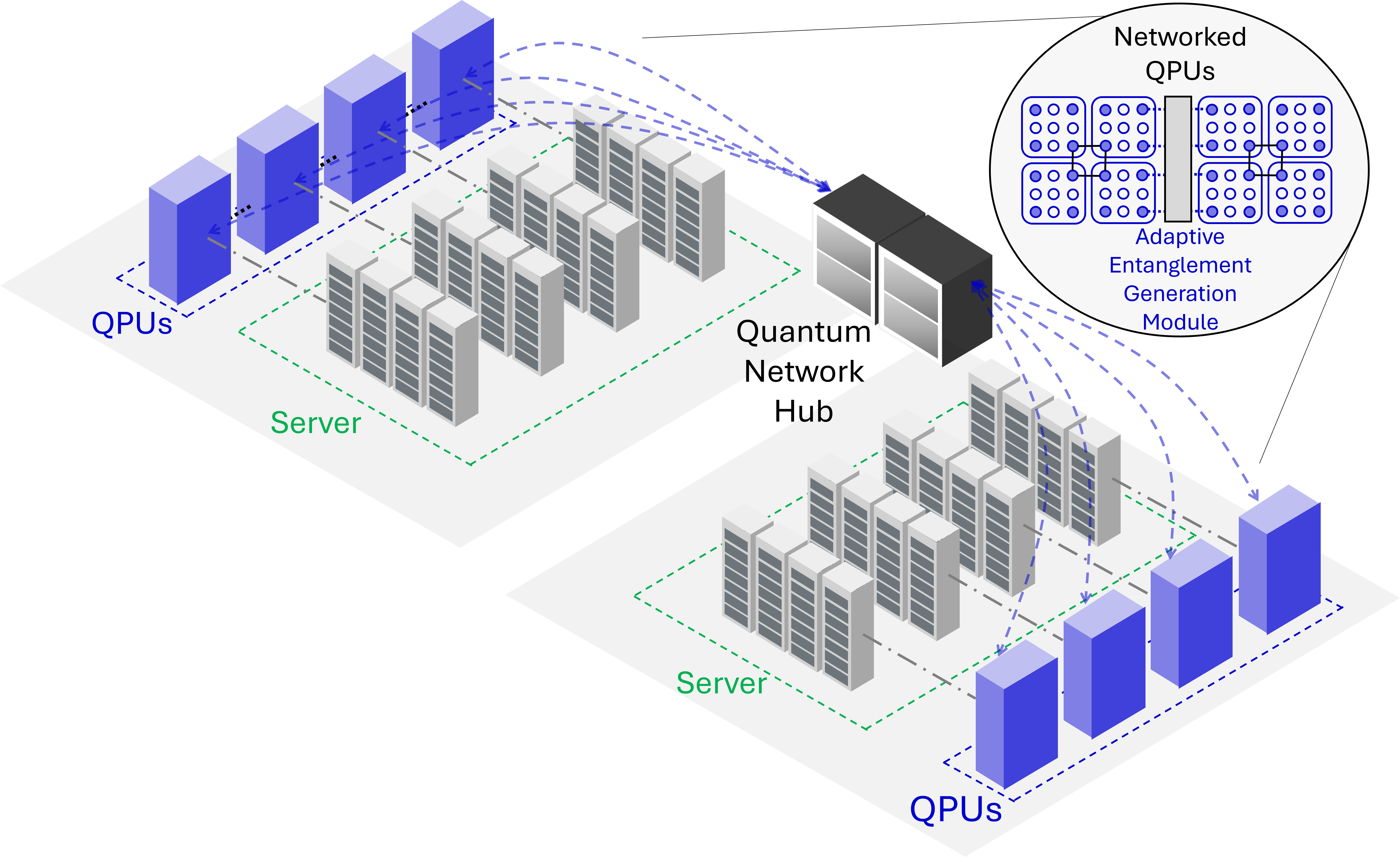}
  \caption{Quantum‑HPC implementation of the ModEn‑Hub architecture. Heterogeneous QPUs and classical control servers are interconnected via a central Quantum Network Hub that delivers on‑demand Bell pairs and orchestrates teleportation‑based non‑local gates. The inset shows a reconfigurable, blocking circuit‑switched photonic fabric of bounded degree (not full mesh): any‑to‑any connections are scheduled on demand using time/frequency multiplexing and, when needed, entanglement swapping, subject to hub/QPU port constraints.}
  \label{fig:dqc-concept}
\end{figure*}

At the centre of the proposed  ModEn‑Hub (shown in Fig.~\ref{fig:dqc-concept}) is the \emph{Quantum Network Hub}, a smart switching node that delivers on-demand entanglement while avoiding the hardware explosion that would arise if every QPU were directly wired to every other. Each individual processor, or small processor cluster, connects to this hub via a photonic link, allowing the hub to mediate entanglement among any pair of nodes as required.

A key component inside the hub is the Entanglement Generation Module. Acting as a programmable source, it produces Bell pairs and routes each photon to a designated output port connected to a target QPU. This module integrates tunable entanglement sources—such as spontaneous parametric down-conversion or quantum-dot-based emitters—with Bell-state measurement stations to enable entanglement swapping. Real-time orchestration allows dynamic adaptation: if a particular fibre link shows reduced heralded success probability, the module can reroute entanglement attempts through alternate paths or initiate multiple parallel trials until successful distribution is confirmed. Ongoing experimental advancements continue to improve the efficiency of photonic entanglement generation and transduction across diverse quantum platforms, including optical, microwave, and telecom domains\cite{rueda2019electro}.

The \emph{Quantum Network Orchestrator} is a classical control unit—possibly a small cluster—that maintains a global map of entanglement, link status, and queued quantum operations. Given a high-level circuit that spans many processors, the orchestrator decomposes it into local gates and inter-processor operations, schedules entanglement generation or swapping, and coordinates the classical messages that complete teleportation-based gates\cite{mengoni2025efficient}. Its role mirrors that of a distributed job scheduler in a conventional cluster, but with the added burden of tracking fragile quantum states and enforcing timing constraints that are tight compared with qubit coherence times.


When a multi-QPU program arrives, the orchestrator first compiles it into QPU-specific sub-tasks and then drafts an \emph{entanglement connection plan}. Suppose processor~A must execute a CNOT whose control qubit resides on A and whose target qubit sits on processor~B. If no Bell pair is cached between the two, the orchestrator triggers the entanglement module to create one: a photon pair is emitted, with one photon routed to A’s receiver and the other to B’s. Upon heralded confirmation, the orchestrator notifies both QPUs that an \emph{e-bit} is available. The processors then perform a teleportation-based CNOT, exchanging measurement results via the classical channel under orchestrator supervision. Measurement outcomes—classical bits—are relayed back to the orchestrator, which updates its global state and issues subsequent commands. Throughout execution, the orchestrator refreshes or reallocates Bell pairs as demand evolves, enabling flexible all-to-all connectivity even though only a single photonic link per QPU is physically installed.

By funnelling traffic through a hub, the architecture reduces hardware overhead from \(\mathcal{O}(N^{2})\) point-to-point links to \(\mathcal{O}(N)\) links for \(N\) processors, while allowing the hub’s optical-switch fabric to scale incrementally—additional ports, sources, and memories can be added as new QPUs join the network. Central management also enables strategies such as maintaining an \emph{entanglement cache}: during idle periods the module pre-generates Bell pairs between frequently interacting nodes, so that subsequent teleportation steps can proceed without the latency of fresh pair creation.

A dedicated low-latency classical channel accompanies every quantum link. Precise timing is essential: operations on spatially separated qubits must be completed within their coherence windows. Advanced synchronisation protocols over optical fibre are being integrated into prototype hubs to ensure temporal alignment. With control latencies on the order of hundreds of nanoseconds across data-centre-scale distances, classical communication becomes effectively negligible compared to typical quantum gate times, allowing the orchestrator to meet stringent feedback requirements.


A near-term prototype could comprise a small number of superconducting or trapped-ion QPUs co-located within a single facility and interconnected via optical fibre to a centralised photonic entanglement module. The proposed architecture emphasises a clear separation between the quantum-optical data plane and the classical control plane, established through well-defined interfaces. This modular design allows both layers to evolve independently—for example, by integrating higher-rate entanglement sources or implementing more sophisticated control algorithms—mirroring the long-standing separation of data and control planes in classical networking systems.

\section{Applications and Use Cases}
A fully realised Modular Entanglement Hub promises to extend quantum computing resources on demand, enabling workloads that would otherwise exceed the capacity of individual processors. The paragraphs below outline representative scenarios in which the proposed architecture is especially impactful.

Modular quantum cloud services can emerge once multiple QPUs—possibly from different vendors and based on heterogeneous technologies—are interconnected in a data-centre environment. Consider a facility that houses several \(100\)-qubit superconducting processors alongside a set of \(50\)-qubit ion-trap devices. A user requesting \(200\) logical qubits submits a job to the cloud interface; the network orchestrator partitions the circuit across three QPUs and employs entanglement links to execute cross-processor gates\cite{burt2025entanglement,mengoni2025efficient}. From the user’s perspective the system behaves as a monolithic \(200\)-qubit computer, yet the underlying hardware remains modular and incrementally upgradable: additional QPUs may be added without replacing existing nodes, provided that entanglement links can be established with acceptable fidelity and rate.

Beyond computation, entangled networks of quantum devices can enhance \emph{distributed sensing and metrology}. In a Quantum-HPC center with distributed quantum computers, certain QPUs may be dedicated to precision magnetic field sensing or optical clock stabilization, while others perform computation. Periodic entanglement distribution allows the orchestrator to form a composite sensor network whose joint measurements outperform any single device. Entangled atomic-clock QPUs, for instance, could constitute a globally synchronised time reference with accuracy exceeding that of individual clocks, coordinated through scheduled entanglement exchange.

Multi-node connectivity also enables \emph{secure multi-party quantum computation}. Organisations that each operate a small quantum processor can jointly evaluate an algorithm without disclosing their private data, relying on protocols related to multi-party key distribution or blind quantum computation\cite{wei2025universal}. The hub facilitates Bell-pair distribution while quantum-encryption techniques ensure that it remains oblivious to the plaintext data. Recent proposals for quantum-internet exchanges envisage such trusted hubs as a central element of cryptographic infrastructure.

Fault-tolerant quantum computing benefits significantly from networked modules. Large-distance error-correcting codes demand sizeable physical-qubit overheads if implemented within a single monolithic device. An alternative is to realise each logical qubit, or small cluster of logical qubits, on a separate module and to connect these modules via high-quality entanglement. The orchestrator enables the dynamic distribution of entanglement between logical blocks, a capability that could support higher-level paradigms. Theoretical studies show that moderate-fidelity interconnects suffice to improve overall fault-tolerance thresholds when arranged in suitable graph states. 


\section{Challenges and Open Issues}

\begin{table}[!b]
  \centering
  \caption{Principal challenges facing Modular Entanglement Hubs and the associated research priorities.}
  \label{tab:challenges}
  \renewcommand{\arraystretch}{1.1}
  \begin{tabular}{p{0.32\linewidth}p{0.63\linewidth}}
    \toprule
    \textbf{Challenge / Open Issue} &
      \textbf{Key Obstacles and Research Directions} \\
    \midrule
    High-fidelity transduction &
      Cryogenic-to-optical and stationary-to-photonic converters must exceed \(>99\,\%\) fidelity while remaining compact, power efficient, and rack-mountable. Integration with superconducting, trapped-ion, and neutral-atom QPUs is still an engineering frontier. \\[0.4em]

    Entanglement distribution at scale &
      Multi-tier topologies require dynamic entanglement routing and congestion control that meet stringent fidelity-latency targets across many hops. Algorithms for adaptive path selection and hierarchical swap scheduling are still immature. \\[0.4em]

    Timing and synchronisation &
      Sub-nanosecond clock alignment is needed for interference-based Bell measurements. Scaling White~Rabbit-class timing to hundreds of QPUs in production data centres raises deployment cost and stability challenges. \\[0.4em]

    Standards, APIs, and programming models &
      The lack of common abstractions for entanglement establishment, error handling, and resource allocation inhibits multi-vendor interoperability. Formal language constructs for circuit partitioning and coordination across disjoint QPUs are still emerging. \\[0.4em]

    Error management and fault tolerance &
      Coordinating error-correction cycles over lossy links introduces latency overhead and new failure modes. Practical schemes must balance physical-qubit overhead, network bandwidth, and achievable logical error rates. \\[0.4em]

    Performance optimisation &
      Joint scheduling of circuit mapping and entanglement generation is largely unexplored. Machine-learning-based resource-management agents offer promise but lack rigorous performance guarantees on real hardware. \\
    \bottomrule
  \end{tabular}
\end{table}

Although the proposed Modular Entanglement Hub (ModEn‑Hub) architecture offers a credible path toward scalable distributed quantum computation, several technical and scientific challenges must be addressed before deployments can reach data‑centre scale. In particular, the data‑centre realisation depicted in the preceding figure—where QPUs and classical servers interface with a central Quantum Network Hub—places stringent requirements on photonic interfaces, network‑level switching, synchronisation, software interoperability, and control‑plane optimisation.

Superconducting QPUs operate at microwave frequencies and therefore require efficient cryogenic‑to‑optical (microwave‑to‑photon) transduction to communicate over photonic links, whereas trapped‑ion and neutral‑atom platforms can emit telecom photons directly but still demand high‑efficiency collection optics. Laboratory demonstrations have exceeded 99\,\% entanglement fidelity for stationary‑to‑photonic qubit conversion, and substantial effort targets microwave‑to‑telecom transduction. Integrating these components into reliable, rack‑mountable modules—consistent with the hub‑centric, data‑centre form factor shown previously—remains an active engineering frontier, and present transducers continue to bound system efficiency.

While a single hub simplifies topology for modest node counts, scaling to hundreds of QPUs will likely necessitate multi‑tier architectures analogous to classical spine–leaf networks. Hierarchical entanglement switching introduces the entanglement‑routing problem, wherein the controller must select multi‑hop swap paths that satisfy end‑to‑end fidelity and latency targets~\cite{abane2025entanglement}. Preliminary algorithms exist, but efficient and robust techniques under dynamic traffic remain limited. Generalising the current design to a network of ModEn‑Hubs raises further questions about inter‑hub synchronisation, long‑haul delays, and fidelity preservation over many hops—issues that are directly implicated by the photonic fan‑out and reconfigurable fabric suggested in the figure.

Because interference‑based Bell measurements require sub‑nanosecond alignment, clock drift or fibre‑length fluctuations can abort entanglement attempts. Precision‑timing solutions such as White Rabbit achieve sub‑nanosecond accuracy over optical fibre and are being integrated into prototypes; however, scaling such infrastructure to hundreds of QPUs and multiple hubs in a production data‑centre entails non‑trivial deployment and stability challenges. These constraints directly underpin the low‑latency classical control channels and real‑time feed‑forward highlighted in the figure.

Despite the execution model provided by the orchestrator, programmers lack mature language constructs to express parallelism across disjoint QPUs and to bind network constraints into compilation. Protocols for entanglement establishment and confirmation, error handling, and resource allocation must be formalised to enable multi‑vendor interoperability. Standards initiatives are beginning to address these needs, yet a plug‑and‑play ecosystem remains distant. Security considerations for multi‑tenant facilities—particularly workload isolation and authenticated entanglement—are likewise insufficiently specified for the data‑centre setting.

Distributed execution introduces additional failure points, since every entanglement link or swap operation can corrupt qubits. Full quantum error correction (e.g.,~\cite{kim2024fault}) appears unavoidable for large‑scale applications, but coordinating error‑correction cycles across networked QPUs remains open. One approach retains only logical (error‑corrected) qubits at each node and teleports data among them, at the cost of increased physical‑qubit overhead and coupling of network latency to code‑cycle duration. Near‑term systems may instead rely on lighter error‑mitigation strategies, accepting limited circuit depth in exchange for reduced complexity.

Scheduling decisions—how to map circuits onto heterogeneous QPUs, when to trigger entanglement generation or swapping, and how to leverage cached Bell pairs—strongly affect throughput and latency. While heuristics from classical networking and recent quantum‑routing work inform current orchestrators, richer optimisation methods (including machine‑learning‑based resource‑management agents) are likely to yield substantial gains. As prototypes mature, a dedicated subfield of quantum‑network resource management is expected to emerge, directly aligned with the orchestration roles indicated in the data‑centre deployment figure.

Despite these open problems, rapid advances in photonic interfaces, precision timing, and quantum‑network prototypes indicate steady progress. Converging academic and industrial efforts on transduction, routing, synchronisation, and standardisation will accelerate the viability of hub‑centric architectures such as ModEn‑Hub, bringing scalable, data‑centre‑style networked quantum computing closer to practice.

\section{Results}

\begin{figure}[!b]
  \centering
  \includegraphics[width=0.45\textwidth]{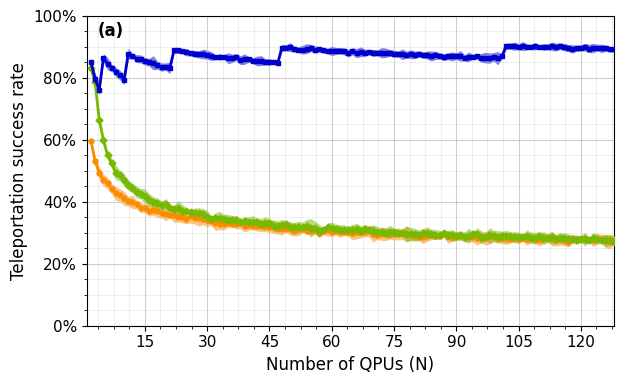}\vspace{-0.5ex}
  \includegraphics[width=0.45\textwidth]{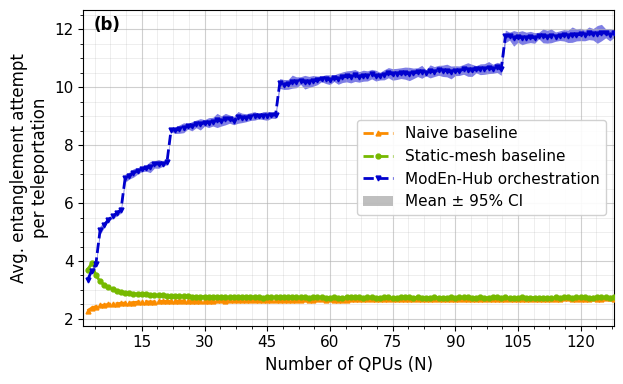}\vspace{-0.5ex}
  \caption{Scalability and cost of distributed teleportation under two strategies. 
(a) Teleportation success rate versus number of QPUs ($N$): the orchestrated policy sustains high performance as $N$ grows, while the naïve sequential baseline degrades. 
(b) Average entanglement attempts per teleportation, showing the higher resource expenditure required by orchestration to obtain the reliability gains in (a).}
  \label{fig:scalability}
\end{figure}


We evaluate distributed teleportation over a growing multi--QPU network using a lightweight Monte Carlo simulaton. For each network size $N\in[1,128]$, we draw $2{,}500$ independent source--destination pairs $(s,d)$ uniformly at random with $s\neq d$. A \emph{round} is one entanglement–generation window during which a node pair may attempt to create a heralded Bell pair. The naïve baseline performs one attempt per round (no reuse); the orchestrated policy performs $K(N)$ parallel attempts per round and caches at most one additional successful Bell pair per unordered node pair. We impose a tight round budget of $R{=}3$; if no Bell pair is obtained within $R$ rounds, teleportation is declared a failure for that request. The per–attempt success probability models aggregate physical loss and control overhead and degrades with scale as \(p_{\mathrm{eff}}(N)=\frac{p_0}{1+\beta\log_2 N}\!\) ,

where $p_0=0.35$ captures small--network transmissivity and $\beta=0.35$ captures increasing attenuation and contention (e.g., longer optical paths, coupling/detection loss, time–multiplexing and classical timing overhead). Under orchestration, parallelism scales logarithmically with size, $K(N)=\max\{2,\lceil\kappa\log_2 N\rceil\}$ with $\kappa=0.9$, giving one–round success probability $p_{\mathrm{round}}(N)=1-(1-p_{\mathrm{eff}}(N))^{K(N)}$. Over $R$ rounds the success event is $1-\bigl(1-p_{\mathrm{round}}(N)\bigr)^R$. Caching is opportunistic: if multiple parallel attempts succeed in the same round, one spare pair is stored (capacity~$=1$) and may serve a future request for that node pair with zero additional attempts and zero waiting.

For each $(N,\text{strategy})$ we report: (i) the \emph{teleportation success rate}, i.e., $\Pr[\text{at least one Bell pair within }R\text{ rounds}]$, and (ii) the \emph{average entanglement attempts per teleportation} (counting all parallel attempts; cache hits contribute $0$). Loss mechanisms (source brightness, coupling, channel attenuation, detector efficiency, heralding) are aggregated into $p_{\mathrm{eff}}(N)$, which decreases with $N$ to reflect longer paths and multiplexing overhead. With $K(N)=\max\{2,\lceil\kappa\log_2 N\rceil\}$, orchestration converts added parallelism into higher one‑round success $p_{\mathrm{round}}(N)=1-(1-p_{\mathrm{eff}}(N))^{K(N)}$ and thus higher end‑to‑end success under fixed $R$, whereas the single‑attempt baseline degrades with scale. Each point averages $2{,}500$ trials; worst‑case standard error is $\leq\!1\%$ (at $p=0.5$), so error bars are omitted.

Fig.~\ref{fig:scalability}(a) shows that the orchestrated policy sustains high performance across scale (near $\sim\!90\%$ over most $N$), whereas the naïve baseline declines markedly as $N$ grows (to roughly $\sim\!30\%$ at large $N$), demonstrating the advantage of adaptive parallelism and caching under strict round limits. Fig.~\ref{fig:scalability}(b) reports the corresponding cost: the orchestrated method consumes more entanglement attempts on average, increasing with $N$ (about $10$--$12$ attempts at the largest sizes), while the naïve approach remains near a constant $\sim\!3$ attempts. 

Together, Fig.~\ref{fig:scalability}(a) and (b) highlight a clear scalability–efficiency trade‑off: the orchestrator trades additional Bell‑pair generation for higher and more stable end‑to‑end success.

\section{Conclusion}
\label{sec:conclusion}
Here we have presented ModEn‑Hub, a modular architecture that centralizes high‑fidelity entanglement generation in a dedicated hub while distributing quantum computation across peripheral QPUs under a real‑time software orchestrator. By cleanly separating the quantum‑optical data plane from the classical control plane, ModEn‑Hub translates principles from cloud/HPC interconnects to quantum systems while addressing decoherence, probabilistic link formation, and the no‑cloning constraint. Early prototypes suggest that photonic interfaces and low‑latency control can be integrated into compact, rack‑mountable modules, pointing toward progressive scaling from small multi‑QPU ensembles to larger clustered deployments. Future work will quantify and optimize memory–performance trade‑offs for entanglement caching (including decoherence of stored Bell pairs, buffer‑capacity limits, and cache‑eviction policies), integrate orchestration with fault‑tolerant protocols, and extend the framework to heterogeneous workloads such as hybrid sensing and distributed cluster‑state generation.

\section*{Acknowledgments}
This work was supported by the Engineering and Physical Sciences Research Council (EPSRC) under grant number EP/W032643/1.

\bibliographystyle{ieeetr}
\bibliography{references}

\section*{Biographies}

\textbf{Kuan-Cheng Chen} (kuan-cheng.chen17@imperial.ac.uk) received the M.Sc. and Ph.D. degrees from Imperial College London (ICL), London, UK, where his research focused on Quantum Information Processing. He is currently a postdoctoral research associate working on distributed quantum computing in the Department of Electrical and Electronic Engineering, ICL. His research interests include distributed quantum computing systems, quantum-based consensus protocols, and distributed quantum algorithms.

\textbf{Felix Burt} (f.burt23@imperial.ac.uk) is a Ph.D. candidate at Imperial College London, researching algorithms, compilers and protocols for quantum information processing systems. He received the M.Sci. degree in Physics and Philosophy from the University of Bristol in 2022, researching noise in quantum thermal machines. Felix's current work is focused on distributed quantum computing architectures, with a major focus on entanglement-efficient compilation of distributed quantum circuits. He also works for the Quantum Engineering Science and Technology (QuEST) centre at Imperial, helping to bring quantum to wider audiences through public engagement.

\textbf{Nitish K. Panigrahy} (npanigrahy@binghamton.edu) received the Ph.D. degree in computer science from the University of Massachusetts Amherst in 2021. He is currently an Assistant Professor with the School of Computing, The State University of New York, Binghamton, NY, USA. His research interests include modeling, optimization, and performance evaluation of quantum and classical information networks.

\textbf{Kin K. Leung} (kin.leung@imperial.ac.uk) received the B.S. degree from The Chinese University of Hong Kong in 1980 and the M.S. and Ph.D. degrees from the University of California, Los Angeles, in 1982 and 1985, respectively. He is currently the Tanaka Chair Professor with the Departments of Electrical and Electronic Engineering and Computing and Co-Director of the School of Convergence Science in Space, Security and Telecoms at Imperial College London. Previously, he held positions as Distinguished Member of Technical Staff at Bell Labs, Lucent Technologies, Technology Consultant at AT\&T Labs, and Member and Distinguished Member of Technical Staff at AT\&T Bell Labs. He is a Fellow of the Royal Academy of Engineering, IEEE, and IET, and a Member of Academia Europaea. His research interests include distributed optimization, machine learning, communication networks, wireless systems, mobile and quantum computing, and stochastic models. He serves on the editorial boards of ACM Computing Surveys and the International Journal of Sensor Networks.

\end{document}